\documentclass{gen-p-l}
\usepackage{epsf}

\begin{document}

\newcommand\beq{\begin{equation}}
\newcommand\eeq{\end{equation}}
\newcommand\bea{\begin{eqnarray}}
\newcommand\eea{\end{eqnarray}}
\newcommand{\rank}{\rm rank}
\def\oupbs{UPBs\ }
\def\pbs{PB's\ }
\newcommand{\ket}[1]{| #1 \rangle}
\newcommand{\bra}[1]{\langle #1 |}
\newcommand{\braket}[2]{\langle #1 | #2 \rangle}
\newcommand{\proj}[1]{| #1\rangle\!\langle #1 |}
\newcommand{\ba}{\begin{array}}
\newcommand{\ea}{\end{array}}
\newtheorem{theo}{Theorem}
\newtheorem{defi}{Definition}
\newtheorem{lem}{Lemma}
\newtheorem{exam}{Example}
\newtheorem{prop}{Property}
\newtheorem{conj}{Conjecture}
\newtheorem{cor}{Corollary}
\newtheorem{propo}{Proposition}

\newcommand{\blankbox}[2]{%
  \parbox{\columnwidth}{\centering
    \setlength{\fboxsep}{0pt}%
    \fbox{\raisebox{0pt}[#2]{\hspace{#1}}}%
  }%
}

\author{David P. DiVincenzo and Barbara M. Terhal}
 
\title{Product Bases in Quantum Information Theory}
 
\address{\vspace*{1.2ex} \hspace*{0.5ex}{IBM T.J. Watson Research
Center, Yorktown Heights, NY 10598, USA}}
 \email{divince@watson.ibm.com, terhal@watson.ibm.com}
\date{\today}

\begin{abstract}
We review the role of product bases in quantum information theory.
We prove two conjectures which were made in \cite{upb2}, 
namely the existence of two sets of bipartite unextendible product bases, 
in arbitrary dimensions, which are based on a tile construction.
We pose some questions related to complete product bases.
\end{abstract}

 \maketitle
 
\section{Introduction}

Quantum information theory is concerned with the applications 
of quantum mechanics in information theory. One of the 
striking features of quantum mechanics is the capacity of 
quantum states to be entangled, that is, a pure state 
$\ket{\psi} \in {\mathcal H}={\mathcal H}_A \otimes {\mathcal H}_B$ 
need not be of the form $\ket{\psi}=\ket{\psi_A}\otimes \ket{\psi_B}$.

Entanglement has turned out to play a role as a resource in 
quantum information theory; sharing an entangled state enables 
parties $A$ and $B$ (Alice and Bob) to transmit quantum information
to each other by merely sending classical bits, via the protocol of 
quantum teleportation \cite{tele}. It is however not only quantum 
entanglement which marks quantum mechanics as a theory with 
fundamentally different features than classical mechanics. The fact 
that quantum mechanics is concerned with noncommutative objects is
reflected in further applications of quantum mechanics in information 
theory. An early example is the idea of quantum cryptography \cite{bb84}, 
which is based on the disturbance versus information gain trade-off in 
quantum states. A more recent development concerns the local distinguishability of sets of mutually orthogonal unentangled (or 'product') states, which we call product bases \cite{qne}. The idea is the following. 

Given is set $S$ of product states $\{\ket{\alpha_i} \otimes
\ket{\beta_i}\}_{i=1}^{|S|}$ in a bipartite Hilbert space ${\mathcal
H}={\mathcal H}_A \otimes {\mathcal H}_B$. The states have the
property that they are mutually orthogonal.  When any of these states
is presented to an Alice and Bob pair with the question which of the
states is given, then the Alice and Bob pair will be able to answer
the question by simply carrying out a quantum measurement
distinguishing the orthogonal states from each other. However, we may
constrain Alice and Bob in their actions in such a way that Alice is
only allowed to operate on space ${\mathcal H}_A$ and Bob on space
${\mathcal H}_B$. Furthermore, we do allow them to (classically)
communicate their results of measurements and other local
actions. Then we ask again; are they able to tell the states in the
product bases apart? In \cite{qne} it was rigorously proved that, for
a certain set of orthogonal product states, the answer is no.

The lack of local distinguishability is tied up with the fact that the
projectors on a set of orthogonal product states need not be locally
commuting.  This feature of nonlocality has been useful in several
contexts: it has permitted an extension of Gleason's theorem to
multipartite systems\cite{wallach}; it is also involved in a class of
product bases which were introduced in \cite{upb1}, the unextendible
product bases (UPB), which we will discuss further here.  Let us give
the definition of an unextendible product basis in a bipartite Hilbert
space (the definition is analogous in multipartite spaces):

\begin{defi}
Let ${\mathcal H}$ be a finite dimensional Hilbert space
of the form ${\mathcal H}_A \otimes {\mathcal H}_B$. A partial product basis is a set ${\rm S}$ of mutually orthonormal pure product states spanning a proper subspace ${\mathcal H}_{\rm S}$ of ${\mathcal H}$. An unextendible product basis is a partial product basis whose complementary subspace ${\mathcal H}_{\rm S}^{\perp}$ contains no product state.
\end{defi}

Unextendible product bases have proved to be extremely rich mathematical 
objects. It was shown in \cite{upb1} and \cite{upb2}, that aside from features
of local indistinguishability, these product bases relate to the phenomenon 
of bound entanglement \cite{pptnodist}. Furthermore, via the connection with 
bound entanglement, it was shown in \cite{terhalposmap} that 
from every unextendible product basis one can construct an indecomposable 
positive linear map. Also, in \cite{lovalon} a graph theoretic construction of unextendible product bases with minimal size in arbitrary dimensions and parties was presented. In all, we believe that it would be highly desirable to develop a systematic theory of product bases, unextendible, uncompletable, or 
complete but `frustrated', see Section \ref{open}.

In this paper we prove two conjectures which were made in \cite{upb2}, 
namely the existence of two sets of bipartite unextendible
product bases, in arbitrary dimensions. In Section \ref{open} 
we address some questions related to complete product bases. 
Let us first recall the definition of these two sets of candidate UPBs.

 
${\bf GenTiles1}$ is a bipartite product basis  in ${\mathcal H}_n \otimes
{\mathcal H}_n$ where $n$ is even, where ${\mathcal H}_n$ denotes a $n$-dimensional
Hilbert space. These states have a tile structure which in the
case of ${\mathcal H}_6 \otimes {\mathcal H}_6$ is shown in Fig. 1a. The general construction
goes as follows: We label a set of $n$ orthonormal states as
$\ket{0},\ldots, \ket{n-1}$.  We define the set of `vertical tile'
states
\begin{eqnarray}
\ket{V_{mk}}=\ket{k} \otimes \ket{\omega_{m,k+1}}
=\ket{k} \otimes
\sum_{j=0}^{n/2-1} \omega^{jm} \ket{j+k+1 \bmod n},\;\;\; \nonumber \\
m=1,\ldots ,n/2-1,\;\;k=0,\ldots,n-1,
\end{eqnarray}
where $\omega=e^{i 4 \pi/n}$. Similarly, we define the set of
`horizontal tile' states:
\begin{equation}
\ket{H_{mk}}=\ket{\omega_{m,k}} \otimes \ket{k},\;\;\; m=1,\ldots ,n/2-1,\;\;k=0,\ldots,n-1.
\end{equation}
Finally we add a `stopper' state
\begin{equation}
\ket{F}=\sum_{i=0}^{n-1} \sum_{j=0}^{n-1}\ket{i} \otimes \ket{j}.
\end{equation}
The stopper state is not depicted in Fig. 1; as a tile it
would cover the whole $6$ by $6$ square. The representation of the set
as an arrangement of tiles informs us about the orthogonalities among
some of its members.

The second set is called {\bf GenTiles2}.  It is a construction made
in dimensions ${\mathcal H}_m \otimes {\mathcal H}_n$ for $n > 3$, $m \geq 3$
and $n\geq m$.  The construction is illustrated in Fig. 2a. The small
tiles which cover two squares are given by 
\begin{equation}
\ket{S_j}=\frac{1}{\sqrt{2}} (\ket{j}-\ket{j+1 \bmod m}) \otimes
\ket{j},\;\;\;0 \leq j \leq m-1.
\label{tileSdef}
\end{equation}
These short tiles are mutually orthogonal on Bob's side.
The long tiles (in general not contiguous) that stretch out in the vertical
direction in Fig. 2a are given by
\begin{eqnarray}
\ket{L_{jk}}=\ket{j} \otimes \frac{1}{\sqrt{n-2}}
\left(\sum_{i=0}^{m-3} \omega^{ik} \ket{i+j+1 \bmod m}+\sum_{i=m-2}^{n-3}\omega^{ik}\ket{i+2}\right),\;\;\; \nonumber \\
0 \leq j \leq m-1,\;\;1 \leq k \leq n-3,
\end{eqnarray}
with $\omega=e^{i \frac{2\pi}{n-2}}$. Lastly we add a `stopper' state
\begin{equation}
\ket{F}=\frac{1}{\sqrt{nm}} \sum_{i=0}^{m-1}\sum_{j=0}^{n-1} \ket{i} \otimes \ket{j}.
\label{tileFdef}
\end{equation}
The total number of states is $mn-2m+1$.

\section{Two Theorems}

\begin{theo}
The set of states {\bf GenTiles1} forms a UPB on ${\mathcal H}_n\otimes {\mathcal H}_n$ for all even $n\ge 4$.
\end{theo}

\begin{proof}
We proceed by assuming that another product state
$\ket{\xi}$ orthogonal to all these exists, and showing that a
contradiction results.  Expanded in the basis above, $\ket{\xi}$ must
have at least one non-zero term, which we will call $\ket{ij}$.  That
is
\begin{equation}
\ket{\xi}=a\ket{ij}+...,
\end{equation}
with $a\neq 0$.
Now, since simultaneous cyclic permutations of
the $A$ and $B$ Hilbert space leaves {\bf GenTiles1} invariant, we can
always relabel things:
\begin{equation}
\ket{\xi}=a\ket{i-j,0}+...,
\end{equation}
where $i-j$ is understood to be mod $n$.  If $i-j<n/2$ this is the form we will
proceed with.  Otherwise, we transform instead to $\ket{\xi}=a\ket{0,j-i}+...$,
then we interchange $A$ and $B$ with a shift of 1 (another symmetry of this
tile set) to obtain $\ket{\xi}=a\ket{j-i-1,0}+...$.  In either case
the state is written as
\begin{equation}
\ket{\xi}=a\ket{s,0}+...
\end{equation}
with $0\leq s\leq n/2-1$.  Now the real work will consist of narrowing
down what the ... can consist of, given that this state must be
orthogonal to all the states in the set {\bf GenTiles1}, and must be a product state.  

In order for $\ket{\xi}$ to be orthogonal to all the $\ket{H_{m0}}$ states,
it must be of the form
\begin{equation}
\ket{\xi}=a\sum_{s=0}^{s=n/2-1}\ket{s,0}+...
\end{equation}
The state $\ket{\xi}$ must have terms other than the ones shown, since otherwise
it could not be orthogonal to $\ket{F}$.  We consider two cases:  The
easy case is

{\em 1. B only has support on $\ket{0}$.} In this case we can write
\begin{equation}
\ket{\xi}=a\sum_{s=0}^{s=n/2-1}\ket{s,0}+b\ket{t,0}+...
\end{equation}
for some $t$, $n/2\leq t<n$, and some $b\neq 0$.  But now $|\langle
V_{1t}|\xi\rangle|=|b|\neq 0$.  So this case is ruled out; the
other case involves considerably more work:

{\em 2. B has support beyond $\ket{0}$.} In this case we can write
$\ket{\xi}$ as
\begin{equation}
\ket{\xi}=a\sum_{s=0}^{s=n/2-1}\ket{s,0}+b\ket{t,r}+...
\end{equation}
for some $r\neq 0$, $b\neq 0$ and $t<n/2$ (if there were only terms
with $t\geq n/2$, $\ket{\xi}$ would not be a product state).  In fact,
since for a product state the $A$ state must be independent of the
result of a projection by $B$ onto his basis, the state must therefore
have the form
\begin{equation}
\ket{\xi}=a\sum_{s=0}^{s=n/2-1}\ket{s,0}+b\sum_{s=0}^{s=n/2-1}\ket{s,r}+...
\end{equation}
Fig. 1a gives a graphical depiction of this state for $r=4$ and $n=6$.
The rest of the argument is easiest to follow using this series of
pictures.  Fig. 1b shows the additional constraints on the state
arising from the orthogonality with the set of states $\ket{V_{m,n/2-1}}$
(if $r$ had been $r<n/2$, we would have used orthogonality with 
$\ket{V_{m,0}}$ to proceed instead).  In symbols, this state is
\begin{equation}
\ket{\xi}=a\sum_{s=0}^{s=n/2-1}\ket{s,0}+b\sum_{s=0}^{s=n/2-1}\ket{s,r}
+b\sum_{q=n/2,q\neq r}^{q=n-1}\ket{n/2-1,q}+...,
\end{equation}
but at this point it is much easier to understand graphically.  Next
we again impose the constraint that the state have a product form;
graphically this requirement can be explained by saying that all the
columns and rows have to be proportional to one another.  Thus we
``fill out the rectangle'' with $b$s as in Fig. 1c.  In symbols, the
state now has the description
\begin{equation}
\ket{\xi}=a\sum_{s=0}^{s=n/2-1}\ket{s,0}+b\sum_{s=0}^{s=n/2-1}
\sum_{q=n/2}^{s=n-1}\ket{s,q}+...
\end{equation}
We now invoke orthogonality with $\ket{V_{m,0}}$ to fill in the
additional $b$s in Fig. 1d.  (We will now dispense with the algebraic
expressions altogether.)  Then again requiring a product state gives
Fig. 1e.  To arrive at Fig. 1f we invoke the orthogonality for
$\ket{H_{mk}}$, $1\leq k\leq n-1$, except $k\neq n/2$.  Requiring the
product form brings us to Fig. 1g.  Now by enforcing orthogonality for
$\ket{V_{m,n/2}}$ we find that $a=b$, so the entire state is
constrained as in Fig. 1h.  But this state is just $\ket{F}$, so it
fails to be orthogonal to all the states.  Therefore, an additional
product state does not exist, and {\bf GenTiles1} is a UPB.
\end{proof}

\begin{figure}[htbf]
\centering
\epsfxsize=5in
\leavevmode\epsfbox{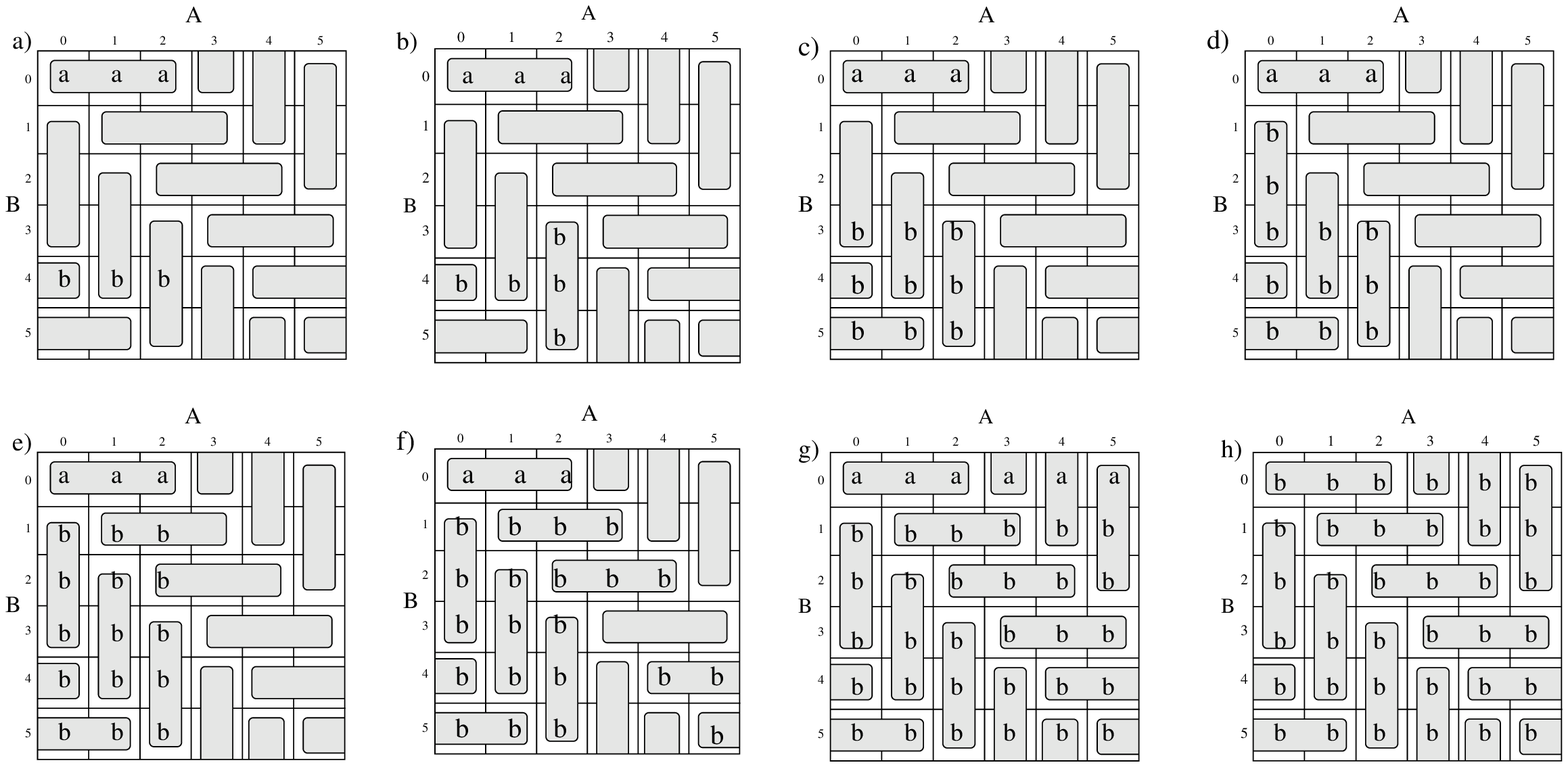}
\caption{}
\end{figure}

\begin{theo}
The set of states {\bf GenTiles2} forms a UPB on ${\mathcal H}_m \otimes
{\mathcal H}_n$ for $n>3$, $m\ge 3$, and $n\ge m$.
\end{theo}

\begin{proof}
We show that this set is a UPB, for all $m$ and $n$, by
the same methods as before.  Since {\bf GenTiles2} has less symmetry
than {\bf GenTiles1} we will have to examine more cases, but the
methods will be the same.  We will number the following paragraphs to
indicate the structure of the cases being considered.

In all cases we begin by assuming that there is an additional product
state $\ket{\xi}$ with nonzero amplitude on some basis state $\ket{ij}$;
we will examine all possible values of $i$ and $j$.  

1. Small-tile case.  Suppose that $i=j=0$, so that the state is
\begin{equation}
\ket{\xi}=a\ket{00}+...  
\end{equation}
Orthogonality with $\ket{S_0}$ gives
\begin{equation}
\ket{\xi}=a(\ket{00}+\ket{10})+...\label{lab1}
\end{equation}
Note that the outcome would have been the same if we had started with
assuming that the amplitude of $\ket{10}$ was nonzero.  By relabeling
of the Hilbert space, this case will cover any initial $\ket{ij}$
which lies in a small tile.
 
1a. Consider the case where the $B$ part of the product state is $\ket{0}$.
Orthogonality with $\ket{F}$ requires that Eq. (\ref{lab1}) have other
nonzero terms:
\begin{equation}
\ket{\xi}=a(\ket{00}+\ket{10})+b\ket{r,0}+...\label{lab2}
\end{equation}
But then $|\langle L_{r1}|\xi\rangle|=|b|\neq 0$, so this case is excluded.

1b. Consider the case where the $B$ part of the product state is not just
$\ket{0}$.  Then we know that
\begin{equation}
\ket{\xi}=a(\ket{00}+\ket{10})+b\ket{r,t}+...\label{lab}
\end{equation}
Here $r$ is 0 or 1, and $t>0$.  We now go down to two other subcases,
depending on whether the additional term is in a small or a large
tile.

\begin{figure}[htbf]
\centering
\epsfxsize=5in
\leavevmode\epsfbox{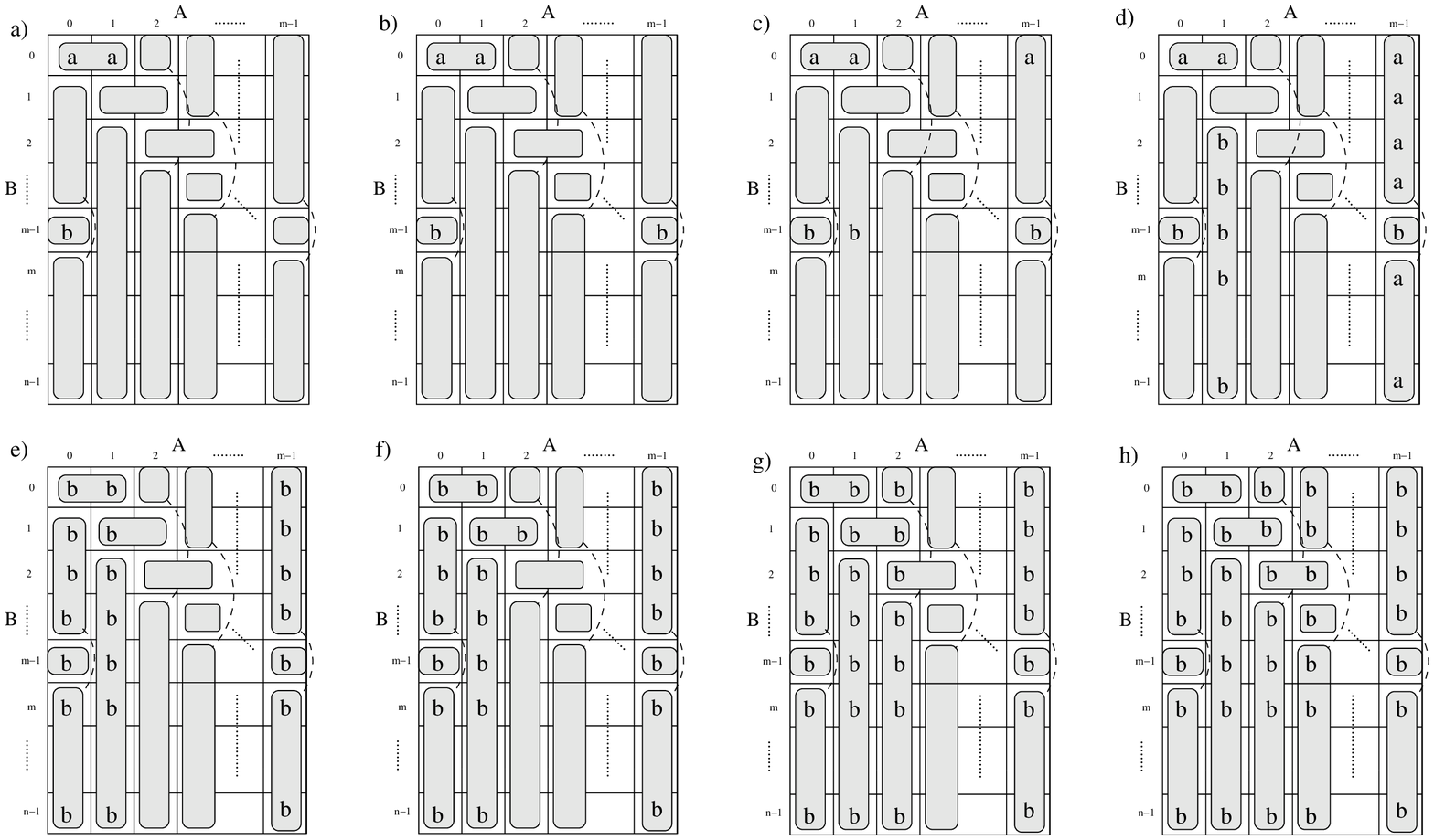}
\caption{}
\end{figure}

1b1. $\ket{r,t}$ is in a small tile (Fig. 2a illustrates the case
$r=0$, $t=m-1$).  Orthogonality with respect to $\ket{S_{m-1}}$
gives Fig. 2b; then the product constraint, which requires rows 0 and
$m-1$ to be proportional in the example shown, gives Fig. 2c.
Orthogonality with respect to $\ket{L_{1k}}$ and $\ket{L_{m-1,k}}$
gives Fig. 2d.  The product state condition, which requires rows $m-2$
and $m-1$ to be proportional, gives $a=b$ and brings us to Fig. 2e.
Orthogonality with $\ket{S_1}$ takes us to Fig. 2f, another
application of the product state condition gives Fig. 2g.  Finally,
repeated application of orthogonality with $\ket{S_i}$, $2\leq i\leq
m-2$, and the product-state condition, fills in the whole state with
$b$s as in Fig. 2h.  But this is just $\ket{F}$, so $\ket{\xi}$ cannot
be orthogonal to all of {\bf GenTiles2} in this case.

1b2. $\ket{r,t}$ is in a large tile.  Fig. 3a shows the case $r=0$,
$t=2$, it will be easy to see that all cases are equivalent.
Orthogonality with $\ket{L_{0k}}$ gives Fig. 3b.  The product state
condition brings us to Fig. 3c.  Orthogonality with $\ket{S_1}$ and
$\ket{L_{1k}}$ gives Fig. 3d, and the product state condition gives
Fig. 3e.  Now, orthogonality with $\ket{L_{2k}}$ requires $a=b$.  The
remainder of the reasoning follows the same track from Fig. 2g; this
case is excluded.

\begin{figure}[htbf]
\centering
\epsfxsize=5in
\leavevmode\epsfbox{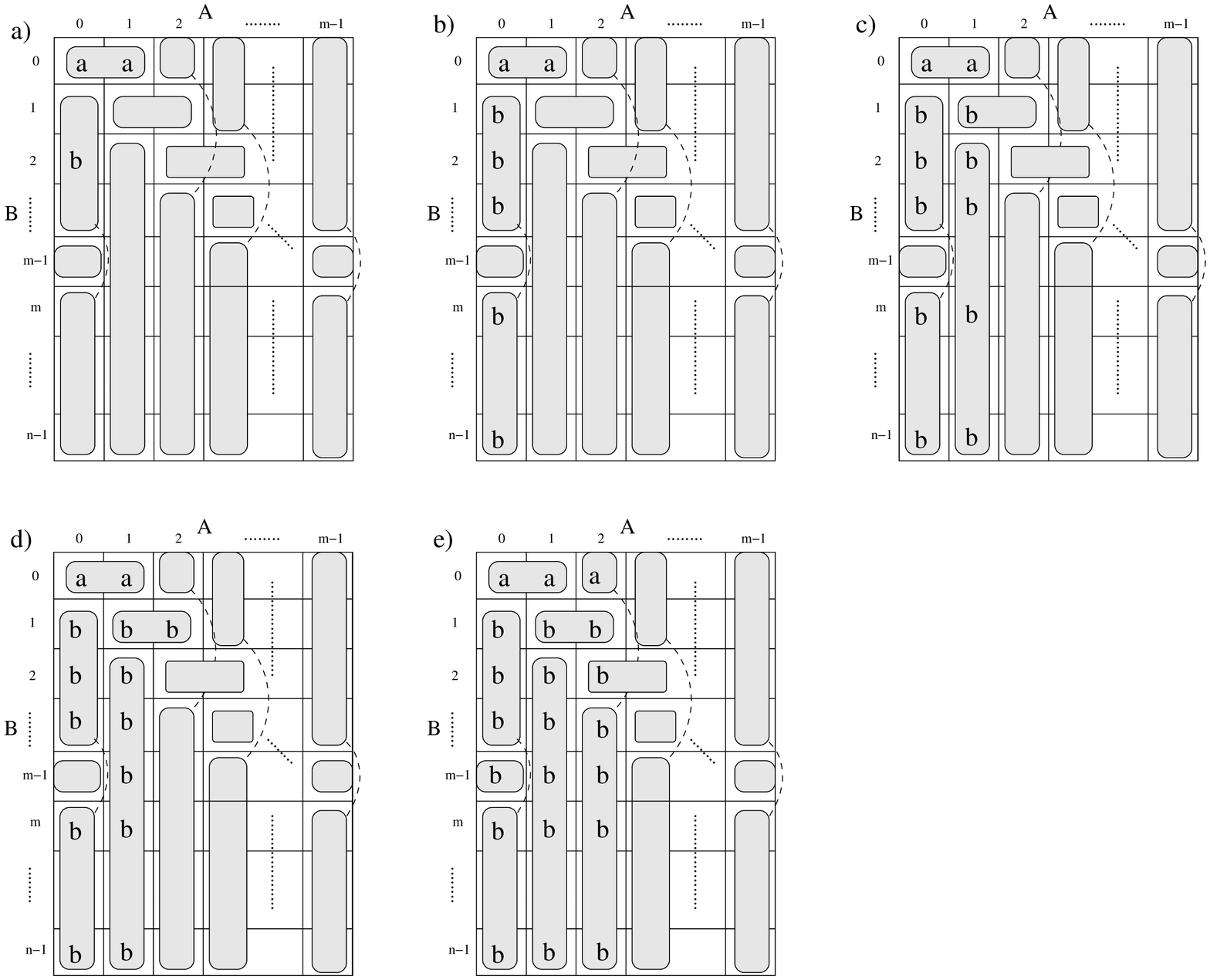}
\caption{}
\end{figure}
 
\begin{figure}[htbf]
\centering
\epsfxsize=5in
\leavevmode\epsfbox{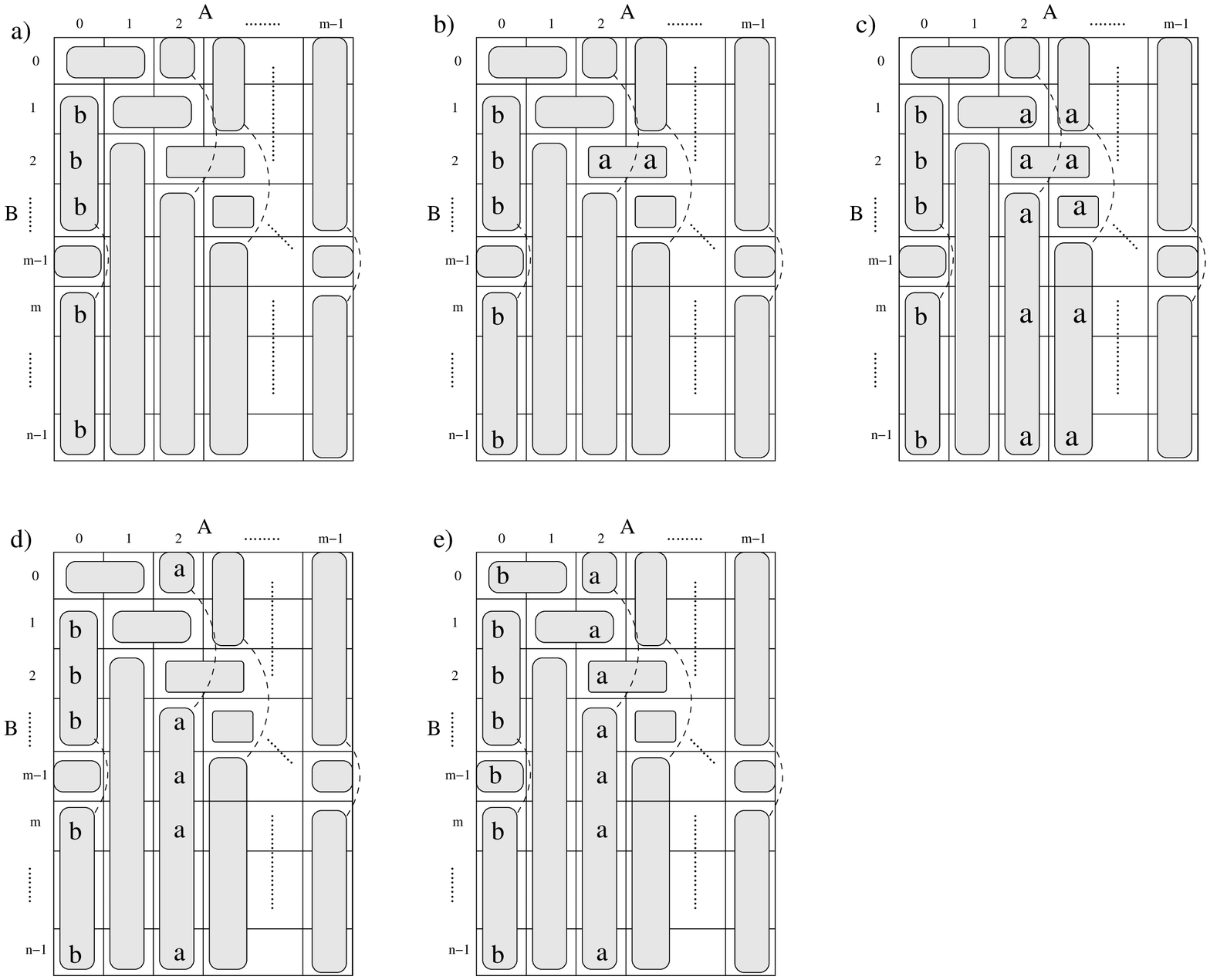}
\caption{}
\end{figure}

2. Large-tile case.  That is, we assume that $\ket{\xi}$ has at least
one non-zero element, and it is in a large tile.  All large tiles
are equivalent by relabeling of the Hilbert space, so we can pick one,
say $i=0$, $j=2$.  Then by orthogonality to $\ket{L_{0k}}$ we get 
Fig. 4a.  There have to be more non-zero entries since $\ket{\xi}$
must be orthogonal to $\ket{F}$.  We consider different subcases
corresponding to which other entry is nonzero.

2a. Component in $\ket{S_0}$ or $\ket{S_{m-1}}$.  This immediately
brings us back to Fig. 3b.
 
2b. Component in another short tile $\ket{S_r}$.  This is illustrated
for $r=2$ in Fig. 4b, using the orthogonality to $\ket{S_r}$.  Invoking
the product state condition gives Fig. 4c; invoking orthogonality to
$\ket{L_{rk}}$ (not shown) brings us back to the situation of Fig. 3b.

2c. Component in another long tile; we illustrate this for $r=2$,
invoking orthogonality to $\ket{L_{rk}}$, in Fig. 4d.  The product
state condition brings us to Fig. 4e; invoking orthogonality to
$\ket{S_0}$ (not shown), we again return to Fig. 3b.

This covers all cases; all methods of constructing the product state
$\ket{\xi}$ lead to a contradiction.  Thus, {\bf GenTiles2} is
a UPB.
\end{proof}

\section{Open questions}
\label{open}


Let us consider some questions related to complete product bases, that 
is, a set of orthonormal product states in some multipartite Hilbert space 
${\mathcal H}$ which span the full space. The simplest basis for a bipartite 
space ${\mathcal H}={\mathcal H}_A \otimes {\mathcal H}_B$ is the 
`Cartesian' basis $\{\ket{i} \otimes \ket{j}\}$, where $\{\ket{i}\}_{i=1}^{\dim {\mathcal H}_A}$ and $\{\ket{j}\}_{j=1}^{\dim {\mathcal H}_B}$ are orthonormal bases for ${\mathcal H}_A$ 
and ${\mathcal H}_B$ resp. The Cartesian basis has the simple property that 
the projectors on the local parts of the product states commute. In 
general one would like to characterize the local noncommutative 
structure, or the level of 'frustration', of a complete product basis; this structure relates to the 
question of whether members of a product basis for example are distinguishable
by local quantum operations and classical communication, see \cite{qne}.
In order to explore this structure we believe that it 
is interesting to introduce the following notion of `winding':

\begin{defi} Let ${\mathcal B}$ be a Cartesian basis for ${\mathcal H}_A \otimes {\mathcal H}_B$. The procedure of winding is the repeated application of the following two steps: \\
1. Choose a subspace ${\mathcal H}_A' \otimes {\mathcal H}_B' \subseteq {\mathcal H}_A \otimes {\mathcal H}_B$ such that for every state $\ket{a} \otimes \ket{b} \in {\mathcal B}$ 
either (1) $\ket{a} \otimes \ket{b} \in {\mathcal H}_A' \otimes {\mathcal H}_B'$ or 
(2) $\ket{a} \otimes \ket{b} \in ({\mathcal H}_A' \otimes {\mathcal H}_B')^{\perp}$. \\
2. Apply a local unitary transformation $U_A \otimes U_B \in B({\mathcal H}_A' \otimes {\mathcal H}_B')$ (only supported on ${\mathcal H}_A' \otimes {\mathcal H}_B'$) on the set ${\mathcal B}$. Call the new set of partially rotated states ${\mathcal B}'$.
\label{defwind}
\end{defi}

We refer to the reverse procedure as `unwinding'.  We note that with
this winding procedure ${\mathcal B} \rightarrow {\mathcal B}' \rightarrow
\ldots \rightarrow {\mathcal B}_{end}$, the states in each ${\mathcal
B}$-set remain orthogonal product states spanning ${\mathcal H}_A
\otimes {\mathcal H}_B$.  The definition is analogous for complete
product bases in multipartite spaces.

By a procedure of winding we can create a complete product basis that
has a certain level of local noncommutativity which may be characterized
by the number of winding moves. The question whether winding is a good 
characterization of the local structure of the basis relates to the 
question whether, say, all bipartite complete product bases can be 
generated in this manner, i.e.

\smallskip
{\em Question: Are all bipartite complete product bases unwindable?}
\smallskip 

We have not been able to answer this question in general, but we do
know that the answer is yes for all complete product bases in
${\mathcal H}_3 \otimes {\mathcal H}_3$, and ${\mathcal H}_2 \otimes
{\mathcal H}_n$ for all $n=2,3,\ldots$.  Furthermore we note that it is possible to prove
that every complete product basis that is distinguishable by local
quantum operations and classical communication \cite{qne} is
unwindable.


Our question becomes more interesting for multipartite spaces.  We
have shown that all bases in ${\mathcal H}_2 \otimes {\mathcal H}_2
\otimes {\mathcal H}_2$ are unwindable.  However, there is a surprise:
there exists an example of a complete product basis in a 10-partite
Hilbert space (${\mathcal H}_2^{\otimes 10}$) that can be proved to be
{\em not unwindable} \cite{shorpriv}.  This basis emerged from work of
Lagarias and Shor \cite{keller1,keller2} that disproved the Keller
conjecture of tiling theory in 10 dimensions.  In fact, it is the
features of this construction that make it a good counterexample to
the Keller conjecture that also make it fail to conform to Definition
\ref{defwind}: the failure of the `face-to-face' tiling property of
Keller corresponds to there being no pair of states confined to a
subspace ${\mathcal H}_2\otimes{\mathcal H}_1^{\otimes 9}$; and the
fact that the Lagarias-Shor construction contains no
smaller-dimensional counterexamples to the Keller conjecture implies
that there are no sets of four states confined to a subspace
${\mathcal H}_2^{\otimes 2}\otimes{\mathcal H}_1^{\otimes 9}$, no sets
of eight states confined to ${\mathcal H}_2^{\otimes
3}\otimes{\mathcal H}_1^{\otimes 8}$, etc.  These are all the possible
subspaces in Definition \ref{defwind}, so no winding moves are possible
for this basis.  Perhaps further remarkable connections will emerge in
the future with tiling theory that will permit further progress to be
made on our Question.

\bibliographystyle{amsalpha}
\bibliography{refs}

\newcommand{\etalchar}[1]{$^{#1}$}
\providecommand{\bysame}{\leavevmode\hbox to3em{\hrulefill}\thinspace}
\begin{thebibliography}{BDM{\etalchar{+}}99}

\bibitem[AL]{lovalon}
N.~Alon and L.~Lov\'asz, \emph{Unextendible product bases}, Manuscript Jan.
  2000.

\bibitem[BB84]{bb84}
C.~H. Bennett and G.~Brassard, \emph{Quantum cryptography: Public key
  distribution and coin tossing}, Proceedings of the IEEE International
  Conference on Computers, Systems and Signal Processing, 1984, pp.~175--179.

\bibitem[BBC{\etalchar{+}}93]{tele}
C.H. Bennett, G.~Brassard, C.~Cr{\'e}peau, R.~Jozsa, A.~Peres, and W.K.
  Wootters, \emph{Teleporting an unknown quantum state via dual classical and
  {Einstein-Podolsky-Rosen} channels}, Physical Review Letters \textbf{70}
  (1993), 1895--1899.

\bibitem[BDF{\etalchar{+}}99]{qne}
C.H. Bennett, D.P. DiVincenzo, C.A. Fuchs, T.~Mor, E.M. Rains, P.W. Shor, J.A.
  Smolin, and W.K. Wootters, \emph{Quantum nonlocality without entanglement},
  Physical Review A \textbf{59} (1999), 1070--1091.

\bibitem[BDM{\etalchar{+}}99]{upb1}
C.H. Bennett, D.P. DiVincenzo, T.~Mor, P.W. Shor, J.A. Smolin, and B.M. Terhal,
  \emph{Unextendible product bases and bound entanglement}, Physical Review
  Letters \textbf{82} (1999), 5385--5388.

\bibitem[DMS{\etalchar{+}}]{upb2}
D.P. DiVincenzo, T.~Mor, P.W. Shor, J.A. Smolin, and B.M. Terhal,
  \emph{Unextendible product bases, uncompletable product bases and bound
  entanglement}, submitted to Comm. Math. Phys., quant-ph/9908070.

\bibitem[HHH98]{pptnodist}
M.~Horodecki, P.~Horodecki, and R.~Horodecki, \emph{Mixed state entanglement
  and distillation: is there a `bound' entanglement in nature?}, Physical
  Review Letters \textbf{80} (1998), 5239--5242.

\bibitem[LS92]{keller1}
J.C. Lagarias and P.W. Shor, \emph{Keller's cube-tiling conjecture is false in
  high dimensions}, Bull. Amer. Math. Soc. \textbf{27} (1992), 279--283.

\bibitem[LS94]{keller2}
J.C. Lagarias and P.W. Shor, \emph{Cube-tilings of {$R^n$} and nonlinear
  codes}, Discrete and Computational Geometry \textbf{11} (1994), 359--391.

\bibitem[Sho]{shorpriv}
P.~Shor, \emph{Private communication}.

\bibitem[Ter]{terhalposmap}
B.M. Terhal, \emph{A family of indecomposable positive linear maps based on
  entangled quantum states}, to appear in Lin. Alg and Its Appl.,
  quant-ph/9810091.

\bibitem[Wal]{wallach}
N.~R. Wallach, \emph{An unentangled {Gleason's} theorem}, quant-ph/0002058.

\end{thebibliography}

\end{document}